\documentclass[twoside]{article}
\usepackage{fleqn,espcrc2}

\usepackage{epsfig}

\def\simge{\mathrel{\rlap{\raise 0.511ex \hbox{$>$}}{\lower 0.511ex 
  \hbox{$\sim$}}}}
\def\simle{\mathrel{\rlap{\raise 0.511ex \hbox{$<$}}{\lower 0.511ex 
  \hbox{$\sim$}}}} 
\def\slash#1{\setbox0=\hbox{$#1$}\dimen0=\wd0 \setbox1=\hbox{/} \dimen1=\wd1
  \ifdim\dimen0>\dimen1 \rlap{\hbox to \dimen0{\hfil/\hfil}} #1  \else 
  \rlap{\hbox to \dimen1{\hfil$#1$\hfil}}  / \fi}

\newcommand{\be}{\begin{equation}}
\newcommand{\ee}{\end{equation}}
\newcommand{\bea}{\begin{eqnarray}}
\newcommand{\eea}{\end{eqnarray}}
\newcommand{\msb}{\overline{\rm{MS}}}
\newcommand{\mev}{\ {\rm MeV}}   
\newcommand{\gev}{\ {\rm GeV}}   

\newcommand{\mud}{\overline{m}_{ud}}

\newcommand{\ms}{\overline{m}_s}
\newcommand{\mb}{\overline{m}_b}
\newcommand{\Oa}{{\cal O}(a)}

\title{Quark Masses on the Lattice: Light and Heavy
\vspace{-3.5cm}
\begin{flushright}
\small RM3-TH/00-21
\end{flushright}
\vspace{2.7cm}}

\author{V. Lubicz\address{Dipartimento di Fisica, Universit\`a di Roma Tre and 
INFN, Sezione di Roma III, \\ Via della Vasca Navale 84, I-00146 Rome, Italy}}
       
\begin{document}

\begin{abstract}
I review the current status of lattice calculations of light and heavy quark 
masses. Significant progresses, in these studies, have been allowed by 
the introduction of improved actions and non-perturbative renormalization 
techniques. Current determinations of light quark masses are accurate at the 
level of 20\%, where the main source of uncertainty is represented by the 
quenching error. The determination of the bottom quark mass is accurate at the 
impressive level of 2\%. As final averages of lattice results, I quote 
$\mud(2\gev) = (4.5 \pm 1.0) \mev$, $\ms(2\gev) = (110 \pm 25) \mev$ and 
$\mb(\mb) = (4.26 \pm 0.09) \gev$.
\vspace{1pc}
\end{abstract}

\maketitle

\section{INTRODUCTION}\label{sect:intro}

Calculations of quark masses are one of the most intensive subject of 
investigation for lattice QCD. An accurate determination of these 
pa\-ra\-me\-ters is in fact extremely important, for both phenomenological and 
theoretical applications. The charm and bottom masses, for instance, enter 
through the heavy quark expansion, the theoretical expressions of several cross 
sections and decay rates. From a theoretical point of view, an accurate 
determination of quark masses may give insights on the physics of flavour, 
revealing relations between masses and mixing angles, or specific textures in 
the quark mass matrix, which may originate from still uncovered flavour 
symmetries.

The values of quark masses cannot be directly measured in the experiments 
because quarks are confined inside the hadrons. On the other hand, quark masses 
are fundamental (free) parameters of the theory and, as such, they cannot be 
computed on the basis of purely theoretical con\-si\-de\-ra\-tions. The values of
quark masses can be only determined by comparing the theoretical evaluation of 
a given physical quantity, which depends on quark masses, with the corresponding
experimental value. Typically the pion, kaon and $\phi$ meson masses are used on
the lattice to compute the values of the light quark masses, whereas the 
$b$-quark mass is determined by computing the mass of the $B$ or the $\Upsilon$ 
mesons. Different choices are all equivalent in principle, and the differences 
in the results, obtained by using different hadron masses as input parameters, 
give an estimate of the systematic error.

As all other parameters of the Standard Model lagrangian, quark masses can be 
defined as effective couplings, which are both renormalization scheme and scale 
dependent. A scheme commonly adopted for quark masses is the $\msb$ scheme, with
a renormalization scale chosen in the short-distance region in order to make
this quantity accessible to perturbative calculations. It is a common practice 
to quote the values of the light quark masses at the renormalization scale 
$\mu=2$ GeV, whereas the heavy quark masses are usually given at the scale of 
the quark mass itself, e.g. $\mb(\mb)$. This convention will be followed in the 
present review. 

At Lattice 99, no plenary talk was dedicated to review lattice calculations of 
quark masses. For this reason, I will mainly report on results presented in the 
last two years. With respect to previous determinations, significant progresses 
in controlling the systematic errors have been allowed by the extensive 
implementation of non-perturbative $\Oa$-improvement~\cite{alpha} and 
non-perturbative renormalization techniques~\cite{npm,npm_sf}. The effect of 
these tools, on lattice calculations of quark masses, will be discussed in 
sec.~\ref{sect:theory}. The recent determinations of light quark masses will be 
presented in sec.~\ref{sect:light}, and of the $b$-quark mass in 
sec.~\ref{sect:heavy}. Some comments on the determination of the charm quark 
mass will be done in sec.~\ref{sect:conclu}, where I will also summarize the 
lattice ``world averages" of quark masses and present my conclusions.

\section{DEFINITIONS AND THEORETICAL PROGRESSES}\label{sect:theory}
\subsection{Quark Masses and ${\cal O}(a)$-improvement}
Non perturbative definitions of quark masses are provided by the chiral Ward
identities (WI) of QCD~\cite{bochicchio}. These definitions allow us to express 
the renormalized quark mass, in a given scheme and at a given renormalization 
scale, in terms of lattice renormalization constants and bare quantities 
(action parameters, matrix elements or correlation functions). Three independent
definitions have been proposed so far which are all equivalent in principle, in 
the sense that they lead to the same value of the renormalized mass. Let us 
discuss these definitions in some details.

\vspace{0.2truecm}\noindent 
{\bf\tt 1) Vector Ward Identity definition:} \\
In terms of renormalized quantities, the chiral vector WI reads:
\be 
\nabla_\mu \langle \alpha \vert \hat V_\mu \vert \beta \rangle =
\left( m_1(\mu)- m_2(\mu) \right) 
\langle \alpha \vert \hat S(\mu) \vert \beta \rangle
\ee
where $\vert \alpha \rangle$ and $\vert \beta \rangle$ are arbitrary external
states. Vector current conservation, within the lattice regularization, implies 
a lattice version of the vector WI, from which the following definition of the 
renormalized quark mass, in terms of the (bare) Wilson parameter, can be 
derived:
\be
m(\mu)= Z_S^{-1}(a\mu) m = Z_S^{-1}(a\mu) \frac{1}{2a} \left(
\frac{1}{K}-\frac{1}{K_c} \right)
\ee
An $\Oa$-improved definition is obtained with the simple 
replacement~\cite{alpha}:
\be
m \to m \left(1+a\, b_m m \right) \quad , \quad (b_m=b_S/2)
\ee

\vspace{0.2truecm}\noindent 
{\bf\tt 2) Axial Ward Identity definition:}\\
In the Wilson formulation of lattice QCD, axial symmetry is explicitly broken. 
The axial WI can be then imposed only for renormalized quantities, and, in the
case of degenerate masses, takes the form:
\be 
\nabla_\mu \langle \alpha \vert \hat A_\mu \vert \beta \rangle =
2\, m(\mu) \, \langle \alpha \vert \hat P(\mu) \vert \beta \rangle
\ee
From this identity, an alternative definition of the quark mass is derived:
\be
m(\mu) = \frac{Z_A}{Z_P(a\mu)} m^{\mbox{\scriptsize{\rm AWI}}} = 
\frac{Z_A}{Z_P(a\mu)} \frac{\nabla_\mu \langle \alpha \vert A_\mu 
\vert \beta \rangle}{2\, \langle \alpha \vert P \vert \beta \rangle} 
\ee
In this case, $\Oa$-improvement is achieved with the replacement~\cite{alpha}:
\bea
m^{\mbox{\scriptsize{\rm AWI}}} &\to&  \left[ 1+a\, (b_A-b_P)\, m \right]
\cdot \nonumber \\ 
&& \cdot \frac{\nabla_\mu \langle \alpha \vert \left( A_\mu +a\, c_A 
\nabla_\mu P\right) \vert \beta \rangle}{2\, \langle \alpha \vert P(\mu)\vert 
\beta\rangle}
\eea
In the absence of explicit chiral symmetry brea\-king induced by the Wilson term
in the action, the vector and axial vector WI definitions are not independent 
anymore. This is the case, for instance, of staggered fermions and of the 
proposed realizations of Ginsparg-Wilson fermions (domain wall and overlap 
fermions).

\vspace{0.2truecm}\noindent 
{\bf\tt 3) Definition from the Quark Propagator:} \\
This relatively new method has been proposed by APE~\cite{mq_ape99}. It is 
based on the study of the quark propagator, computed in a fixed gauge, at large 
momentum. In this limit, an OPE can be derived for the trace of the quark 
propagator which, up to known logarithmic corrections, reads~\cite{opeprop} :
\be 
\mathrm{Tr}\, \hat S(p) \simeq \frac{m}{p^2} + \frac{4 \pi \alpha_s}
{3 p^4} \langle \bar \psi \psi \rangle + {\cal O}(1/p^6)
\label{eq:ope}
\ee
The leading term of this expansion is proportional to the renormalized quark 
mass. The Wilson coefficient of this term is fixed by an axial WI which, in 
RI-MOM scheme, implies~\cite{mq_ape99}: 
\be
m(\mu) = \frac{1}{12} \mathrm{Tr}\, \left[\hat S^{-1}(p,\mu) \right]_{p^2=\mu^2}
\ee
Note that the relevant renormalization constant, in this case, is the quark 
field renormalization constant $Z_q$. Cancellation of leading $\Oa$ effects is 
obtained by improving the quark field. Since this field is a non gauge-invariant
operator, and the quark propagator is an off-shell correlation function, 
improvement involves mixing with both non gauge-invariant operators and 
operators vanishing by the equation of motion. The $\Oa$-improvement of the 
quark field is obtained through the replacement~\cite{qimp}:
\be
q \to \left(1+a b_q m \right) [ 1+a c_q^\prime (\slash D + m_0) +
a c_{\mathrm{NGI}} \slash \partial ] q
\label{eq:qimp}
\ee

%________________________________________________________________
\begin{figure}[ht]
%\vspace*{-.55cm}
\epsfxsize=7.5cm 
\epsfysize=5.0cm 
\epsfbox{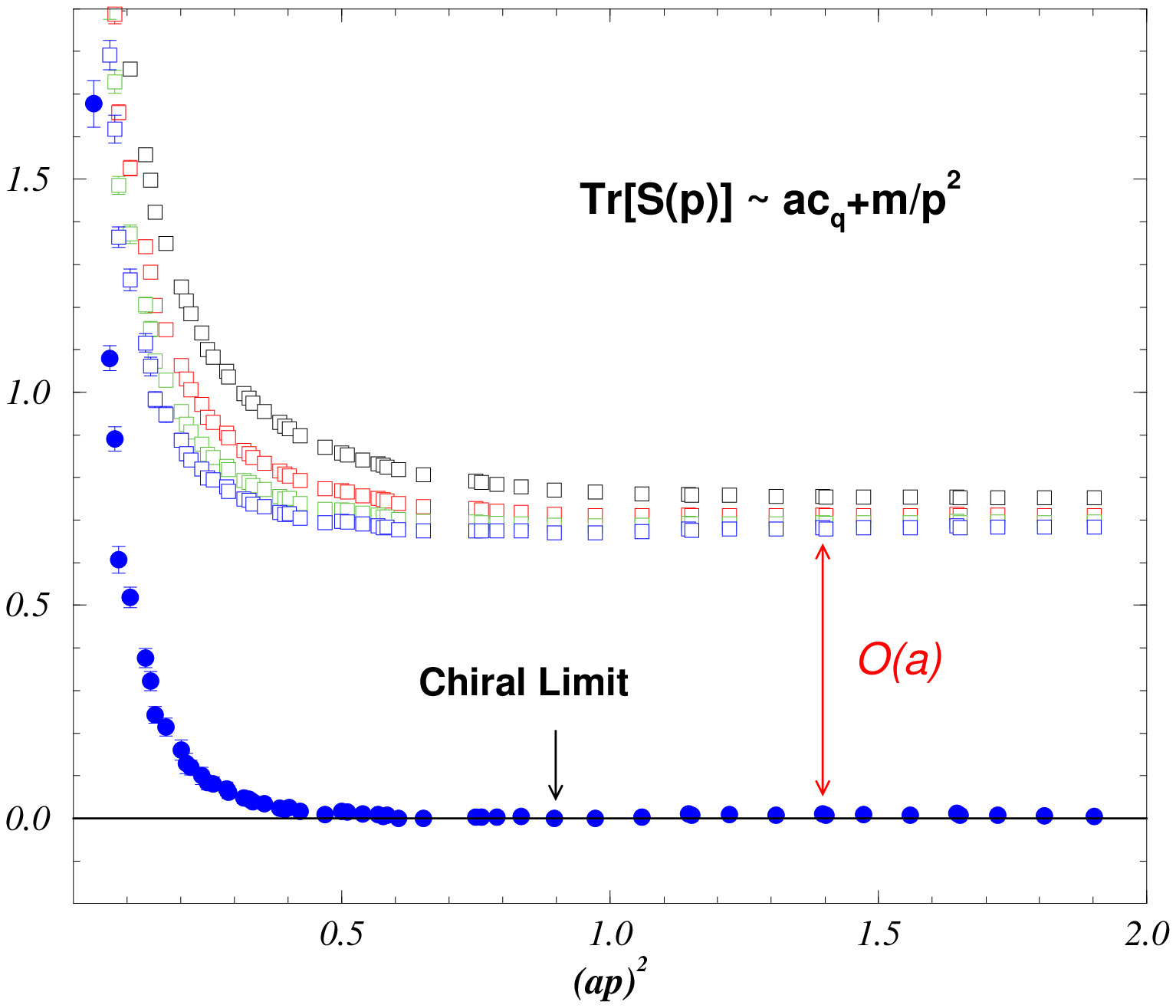}
\epsfxsize=7.5cm 
\epsfysize=4.8cm 
\epsfbox{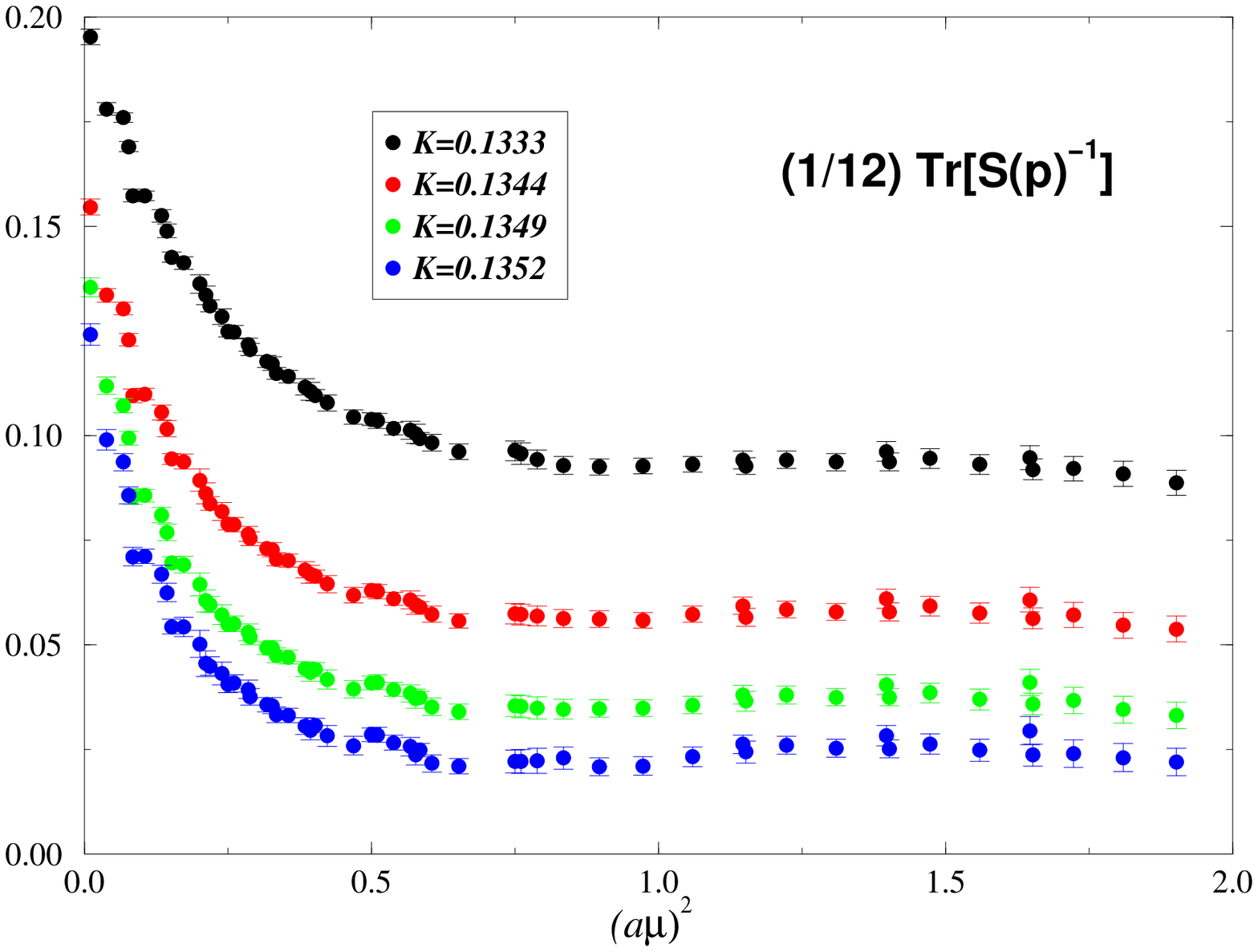}
\vspace*{-1.3cm}
\caption{\it Large momentum behaviour of the lattice quark propagator. The upper 
figure shows the large $\Oa$ term which has to be subtracted before extracting 
the subleading contribution proportional to the quark mass. The behaviour of the
inverse quark propagator, after the subtraction, is shown below. The results 
have been obtained by APE~\cite{mq_ape99} using the non-perturbatively 
$\Oa$-improved Wilson action at $\beta=6.2$.}
\label{fig:mqprop}
\vspace*{-.55cm}
\end{figure}
%________________________________________________________________
In the determination of quark masses with the propagator method, the improvement
procedure is a crucial requirement. The reason is that the OPE of the quark 
propagator on the lattice is dominated by a large lattice artifact. From 
eq.~(\ref{eq:qimp}) one finds, in the large momentum limit and up to small 
logarithmic corrections~\cite{mq_ape99}:
\be 
\frac{1}{12} \mathrm{Tr}\, S(p) \simeq -2 a\,c_q^\prime + 2 a\,c_{\mathrm{NGI}} 
Z_q \equiv a\, c_q
\ee
The lattice quark propagator does not vanish at large $p^2$, as dictated in the 
continuum limit by eq.~(\ref{eq:ope}), but it is dominated by terms of $\Oa$. 
This is illustrated in fig.~\ref{fig:mqprop}. The numerical study of the 
propagator at large $p^2$, in the chiral limit, provides the value of the 
coefficient $c_q$, which can thus be subtracted before computing the subleading 
contribution to the OPE proportional to the quark mass.

The improved calculations of quark masses, with the vector and axial-vector WI 
methods, require the knowledge of several coefficients, $c_A$, $b_A$, 
$b_S$ and $b_P$, besides the coefficient $c_{SW}$ of the clover term in the 
action. Non-perturbative determinations of these coefficients have been 
performed in refs.~\cite{alpha,bb_alpha,bb_lanl}. One finds that the effect of 
improvement, in the determination of quark masses, is quite significant, even 
for light quarks. In the case of the strange quark mass, for instance, an 
estimate of the $\Oa$ effects, with unimproved Wilson fermions, has been
performed by Bhattacharya and Gupta~\cite{gupta} at Lattice 97, based on the
analysis of the continuum extrapolation. They find that discretization errors, 
at two typical values of the lattice coupling, $\beta=6.0$ and $\beta=6.2$, are 
of the order of 25\% and 18\% respectively. These estimates can be compared with 
those obtained with non-perturbatively $\Oa$-improved Wilson fermions, from a 
recent study by ALPHA-UKQCD~\cite{mq_alpha99}. In this case, residual 
discretization effects are reduced at the level of 7\% and 3\% respectively. 
Implementation of $\Oa$-improvement is therefore a crucial ingredient to 
significantly increase the accuracy of lattice calculations of quark masses.

\subsection{Non-perturbative Renormalization}
%___________________________________________________________________________
\begin{table*}
\caption{Values of the strange quark mass, $\ms(2 \gev)$, in MeV, as obtained
by computing the quark mass renormalization constant either in one loop
(boosted) perturbation theory (PT) or using the RI-MOM non-perturbative method 
(NP). The relative difference between the two determinations is also shown.}
\label{tab:npr}
\renewcommand{\tabcolsep}{1.0pc} % enlarge column spacing
\renewcommand{\arraystretch}{1.1} % enlarge line spacing
\begin{tabular}{lccccl}
\hline 
Action & $\beta$& $\ms$ [PT] & $\ms$ [NP] & $\Delta$ &\\  \hline
NP-CLOVER & 6.2 &
$\begin{array}{c} 95(10)^\dag \\ 81(9)^\ddag \end{array}$ & 
$\begin{array}{c} 109(11) \\ 111(13) \end{array}$ & 
$\begin{array}{c} 15\% \\ 37\% \end{array}$ & APE 98 \cite{mq_ape98} \\[10pt]
KOGUT-SUSSKIND & 
$\begin{array}{c} 6.0 \\ 6.2 \\ 6.4 \end{array}$ &
$\begin{array}{c} 80(4) \\ 81(6) \\ 83(4) \end{array}$ &
$\begin{array}{c} 114(5) \\ 110(8) \\ 108(5) \end{array}$ &
$\begin{array}{c} 43\% \\ 36\% \\ 30\% \end{array}$ &
JLQCD 99 \cite{mq_jlqcd99}\\[18pt]
DOMAIN WALL  & $6.0$ & $ 106(8)(14)^\ast$ & $130(11)(18)$ & 23\% & RBC 99 
\cite{mq_rbc99} \\
\hline
\end{tabular}\\[2pt]
$^\dag$ From vector WI; $^\ddag$ from axial WI. $^\ast$ The error is my estimate 
from the RBC data. 
\vspace*{-.3cm}
\end{table*}
%___________________________________________________________________________
Another potential source of systematic errors, in lattice calculations of quark 
masses, is introduced by the truncation of the perturbative expansion in the 
evaluation of the quark mass renormalization constant. Depending on the 
definition considered for the bare mass, this constant is given by $Z_S^{-1}$, 
with the vector WI method, $Z_A/Z_P$, with the axial WI method, or the quark 
field renormalization constant $Z_q$ with the propagator method. In perturbation
theory, all these quantities are known only at one loop. In the last two years,
however, in most of lattice quark masses calculations, the systematic error
associated with the evaluation of the renormalization constant has been reduced 
to a negligible amount by the use of non-perturbative renormalization 
techniques~\cite{npm,npm_sf}. 

Non-perturbative renormalization is reviewed by Sint at this 
conference~\cite{sint}. For this reason,  I will only discuss here those 
particular aspects of this topic which are more closely related to the 
determination of quark masses. 

To show the impact of non-perturbative renormalization in lattice calculations 
of quark masses, I present in table~\ref{tab:npr} the values of the strange 
quark mass, as obtained by computing the quark mass renormalization constant 
either in one-loop (boosted) perturbation theory or with the RI-MOM
non-perturbative method of ref.~\cite{npm}. For illustrative purposes, the 
results of three calculations are considered, which employed three dif\-fe\-rent
fermionic actions: non-perturbative Clover, Kogut-Susskind and domain wall 
fermions. As can be seen from table~\ref{tab:npr}, the difference between 
the results obtained with perturbative and non-perturbative renormalization is 
rather large. Depending on the action, the value of the coupling constant and 
the specific renormalization constant ($Z_S$ or $Z_P$ for Wilson fermions), the 
re\-la\-tive error varies in the range between $\sim 15\%$ and $\sim 40\%$. 
Moreover, at least in the case of staggered fermions, this error decreases as 
one get closer to the continuum limit, but at a very slow rate. Therefore, 
non-perturbative re\-nor\-ma\-li\-za\-tion is required to improve the accuracy of
lattice calculations of quark masses at a significant level (below 15\%). 

I believe that a more detailed discussion deserves the non-perturbative 
calculation of the pseudoscalar renormalization constant, $Z_P$, in the RI-MOM 
scheme. Although rather technical, this discussion is of relevance for the 
determination of quark masses with the axial WI method. Within the 
non-perturbative approach of ref.~\cite{npm}, $Z_P$ is evaluated from the 
amputated correlation function of the pseudoscalar density, computed between 
external off-shell quark states of momentum $p$. This function, $\Gamma_P(p)$,
once conveniently projected onto the matrix $\gamma_5$, is related to the trace 
of the inverse quark propagator by the axial WI: $m\, \widehat \Gamma_P(p) = 
\mathrm{Tr}\, \hat S^{-1}(p)$. The OPE of the quark propagator, given in 
eq.~(\ref{eq:ope}), thus implies:
\be 
\widehat \Gamma_P(p) \simeq C_1(p^2) + C_2(p^2) \, \frac{\langle \bar \psi \psi 
\rangle} {m\, p^2}  + {\cal O}(1/p^4)
\label{eq:gammap}
\ee
The renormalization constant $Z_P$ is determined from the coefficient function
$C_1$. Eq.~(\ref{eq:gammap}) shows that the first power correction, which 
vanishes at large momentum, is however divergent in the chiral limit. 
Physically, this divergence is introduced by the coupling of the pseudoscalar 
density to the pion field, which is the Goldstone boson of the spontaneously 
broken chiral symmetry. It has been emphasized, particularly in 
ref.~\cite{alain}, that, because of the singular behaviour of this term in the 
chiral limit, its contribution to the correlation function $\Gamma_P(p)$ may not
be suppressed enough at large $p^2$, at least for those values of $p^2$ in
actual calculations at which discretization effects are still under control. 
This may affect the determination of $Z_P$ and, according to ref.~\cite{alain},
this systematic error may in fact be the reason of the large discrepancy 
between the perturbative and the non-perturbative determination of $Z_P$. 
Since $Z_m \sim Z_P^{-1}$, the observation of ref.~\cite{alain} is of relevance 
for lattice calculations of quark masses.

\begin{figure}[htb]
%\vspace*{-.55cm}
\epsfxsize=7.5cm 
\epsfysize=5.0cm 
\epsfbox{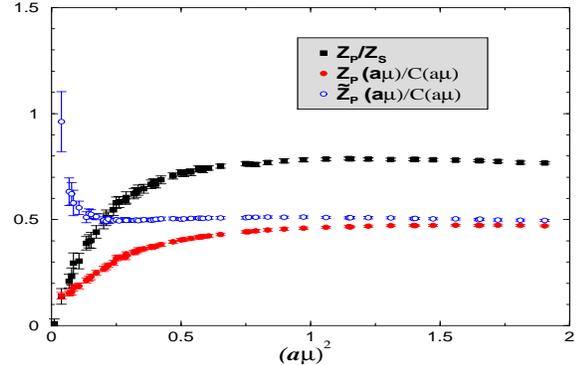}
\vspace*{-1.3cm}
\caption{\it The ratios $Z_P/Z_S$, $Z_P(a\mu)/C(a\mu)$ and $\widetilde 
Z_P(a\mu)/C(a\mu)$ (see text for details) as a function of the scale 
$(a\mu)^2$.}
\label{fig:zp}
\vspace*{-.55cm}
\end{figure}

I believe that the warning of ref.~\cite{alain} deserves some consideration and
a careful investigation. I would like to mention, therefore, three different 
methods which can be used to verify whether the contribution of the Goldstone 
pole, in the actual calculation of $Z_P$, is suppressed at the required level. 
To study this problem, I will use the same APE data which have been used for the
non-perturbative calculation of the renormalization constant of bilinear quark 
operators ($Z_V$, $Z_A$, $Z_S$, $Z_P$ and $Z_T$) with the non-perturbative
Clover action, at $\beta=6.2$~\cite{lubicz}. The first method is based on the 
study of the scale dependence of the ratio $Z_P/Z_S$, computed
non-perturbatively in the RI-MOM scheme. In the large momentum region, where 
power corrections are expected to be negligible, this ratio should exhibit a 
plateau, since the two renormalization constants have equal anomalous 
dimensions. This ratio is shown in fig.~\ref{fig:zp}, and in the large-$p^2$ 
region a plateau is clearly vi\-si\-ble. This suggests that all power 
corrections, including the contribution of the Goldstone pole, are suppressed 
enough. The second method is based on the study of the renormalization group 
invariant combination $Z_P(a\mu)/C(a\mu)$, where the running factor $C(a\mu)$, 
computed in perturbation theory at the NLO, expresses the predicted 
lo\-ga\-ri\-thmic scale dependence of the renormalization constant. Again, in 
the region where non-perturbative power corrections are suppressed enough, the 
ratio $Z_P(a\mu)/C(a\mu)$ should behave as a constant. Indeed, this constant 
behaviour is observed in fig.~\ref{fig:zp}. 
%________________________________________________________________
%________________________________________________________________
Finally, the most efficient way to get rid of the Goldstone boson contribution 
in the calculation of $Z_P$ (but the method can be ea\-si\-ly adapted to the 
calculation of other renormalization constants) has been recently suggested in 
ref.~\cite{giusti}. The idea is to consider a new definition of $Z_P$, which I 
will denote with $\widetilde Z_P$, involving two flavours of non-degenerate 
quarks:
\be
\frac{\widetilde Z_P(\mu)}{Z_q(\mu)} =  
\frac{m_1-m_2}{m_1 \Gamma_P(\mu,m_1) - m_2 \Gamma_P(\mu,m_2)}
\label{eq:zptilde}
\ee
On the r.h.s. of eq.~(\ref{eq:zptilde}), the limit $m_1,m_2\to 0$ is understood.
For $m_2=0$ (and at large va\-lues of the scale), the above definition reduces 
to the standard RI-MOM definition of $Z_P$. However, one immediately realizes 
that, at the leading order in the quark mass, the Goldstone boson contribution 
of eq.~(\ref{eq:gammap}) cancels in the definition of $\widetilde Z_P$, which is
thus well behaved in the chiral limit. To the calculation of $\widetilde Z_P$, 
the warning of ref.~\cite{alain} does not apply anymore. The ratio $\widetilde 
Z_P(\mu)/C(a\mu)$ is plotted in fig.~\ref{fig:zp} as a function of the scale. 
One can see, in this case, that the desired constant behaviour is reached much 
faster than with the ratio $Z_P(\mu)/C(a\mu)$. Moreover, the two ratios converge
to the same value at large scales, within less than 5\%, an uncertainty which is
of the same order of the other systematic errors involved in the calculation. 
This analysis supports the assumption of an adequate suppression of the 
Goldstone pole contribution in previous calculations of $Z_P$.

A significant accuracy in the non-perturbative determination of the quark mass 
renormalization constant has been achieved by ALPHA~\cite{npm_sf}, within the 
Schr\"{o}dinger functional approach. Besides determining the value of mass 
renormalization constant at a low hadronic scale, ALPHA has also computed, 
non-perturbatively, the running factor $C(\mu)$ over a large region extending
from the low hadronic scale up to a very high scale. This calculation, which is 
based on a recursive finite-size scaling technique, allows a fully 
non-perturbative determination of the renormalization group invariant quark 
mass. The non-perturbative calculation of $C(\mu)$ turns out to be in very good 
agreement with the predictions of NLO perturbation theory. The advantage of this
approach is that the matching factor, relating the renormalization group 
invariant mass to the $\msb$ mass, is known at N$^4$LO in perturbation 
theory~\cite{n4lo1,n4lo2}. The remaining perturbative uncertainty in this 
determination is thus reduced at a completely negligible level. The ALPHA
calculation of the mass renormalization constant has been discussed by Sint at 
this conference, and I refer to his review for further details~\cite{sint}.

\section{LIGHT QUARK MASSES}\label{sect:light}

\subsection{Quenched Calculations}
\begin{figure}[htb]
%\vspace*{-.55cm}
\epsfxsize=7.5cm 
\epsfbox{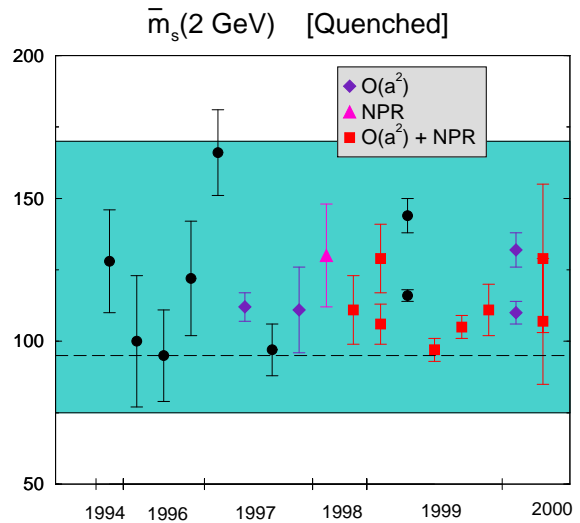}
\vspace*{-1.3cm}
\caption{\it Results of all lattice calculations of the strange quark mass 
obtained, in the quenched approximation, since 1994. The band shows the estimate
of $\ms$ quoted by the particle data group. The dashed line indicates the 
unitarity lower bound on $\ms$.}
\vspace*{-.55cm}
\label{fig:mstrano}
\end{figure}

The determination of light quark masses is, at present, one of the main fields 
of activity of lattice QCD simulations. This is illustrated in 
fig.~\ref{fig:mstrano}, which presents a compilation of all lattice results for 
the strange quark mass, in the quenched approximation, starting from the first 
lattice prediction for $\ms$ obtained with a NLO accuracy~\cite{allton}.
%________________________________________________________________
%________________________________________________________________
These results have been produced by using different actions, different orders of
improvement, different renormalization techniques and dif\-fe\-rent choices of 
the input parameters, used to fix both the scale and the strange quark mass 
itself (either $m_K$ or $m_\phi$). In spite of these differences, we find good 
consistency among the several determinations. In addition, all the results, with
the possible exception of the rather large value obtained by SESAM~\cite{sesam},
suggest an interval of allowed va\-lues for the strange quark mass which is 
smaller than the one quoted by the particle data group (PDG)~\cite{pdg}, 
indicated by the band in the figure. One can also observe that the lattice 
determinations of the strange quark mass cluster close to the unitarity lower 
bound, $\ms \simge 100 \mev$~\cite{srbound}, obtained by using unitarity and 
the positivity of the pseudoscalar spectral function.%
\footnote{The N$^2$LO bound of ref.~\cite{srbound} slightly decreases to
$\ms \simge 95 \mev$, when the N$^3$LO perturbative corrections to the 
pseudoscalar spectral function are taken into account~\cite{damir}.}
Finally, the results which are plotted on the same vertical line in the figure
have been obtained by using either the kaon or the $\phi$ meson mass as input
parameter. It has been shown that a systematic difference of approximately 
$20\%$ is found between the two determinations within the quenched 
approximation.

\begin{figure}[htb]
%\vspace*{-.55cm}
\epsfxsize=7.5cm 
\epsfbox{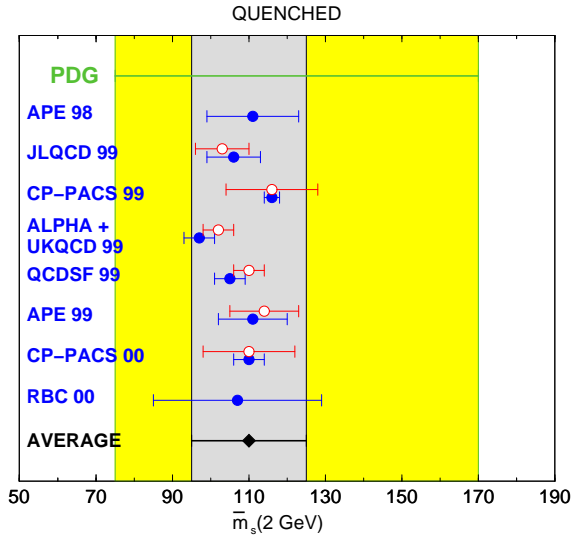}
\vspace*{-1.3cm}
\caption{\it Values of the strange quark mass obtained from recent lattice 
calculations in the quenched approximation (full circles). Empty circles denote 
the values and errors obtained by the author attempting to correct for the 
different systematics.}
%The lowest point (diamond) represents the lattice (quenched) average.}
\vspace*{-.55cm}
\label{fig:msave}
\end{figure}

In order to derive an average value for the strange quark mass, I will 
concentrate in the following on the more recent and accurate results, which are 
given in table~\ref{tab:mslatt} and shown in fig.\ref{fig:msave} with full 
circles.
%___________________________________________________________________________
\begin{table*}[hbt]
\caption{Quenched lattice results for the strange quark mass obtained by using 
in input the value of the kaon mass. Details concerning the action, the value 
of the lattice spacing, the renormalization procedure and the perturbative 
accuracy are also given.}
\label{tab:mslatt}
\renewcommand{\tabcolsep}{0.45pc} % enlarge column spacing
\begin{tabular}{lcccccc}
\hline  
& Action & \multicolumn{2}{c}{$a^{-1}$} & $Z_m$ & PT & $\ms(2 \gev)/\mev$ 
\\  \hline
APE 98 \cite{mq_ape98}         & NP-Clover  & $\sim 2.7 \gev$ & $(m_\rho)$ 
                               & NP-RI & N$^2$LO &  $111 \pm 12$ \\
JLQCD 99 \cite{mq_jlqcd99}     &    KS      &   $a \to 0 $    & $(m_\rho)$ 
                               & NP-RI & N$^2$LO &  $106 \pm 7 $ \\
CP-PACS 99 \cite{mq_cppacs99}  &  Wilson    &   $a \to 0 $    & $(m_\rho)$ 
                               &  PT   & NLO     &  $116 \pm 3 $ \\
ALPHA-UKQCD 99 \cite{mq_alpha99}&  NP-Clover &   $a \to 0 $    & $(r_0,f_K)$
                               & NP-SF & N$^4$LO &  $97 \pm 4 $ \\
QCDSF 99 \cite{mq_qcdsf99}     &  NP-Clover &   $a \to 0 $    & $(r_0)$
                               & NP-SF & N$^4$LO &  $105 \pm 4 $ \\
APE 99 \cite{mq_ape99}         &  NP-Clover & $\sim 2.7 \gev$ & $(m_\rho)$
                               & NP-RI & N$^3$LO &  $111 \pm 9 $ \\
CP-PACS 00 \cite{mq_cppacs00}  &  MF-Clover &   $a \to 0 $    & $(m_\rho)$
                               &   PT  &   NLO   &  $110 ^{+3}_{-4}$ \\
RBC 00 \cite{mq_rbc00}         &    DWF     & $\sim 1.9 \gev$ & $(m_\rho)$
                               & NP-RI & N$^3$LO &  $\begin{array}{c} 
110 \pm 2 \pm 22\ ^\dag \\ 105 \pm 6 \pm 21\ ^\ddag \end{array}$ \\
\hline
\end{tabular}\\[2pt]
$^\dag$ From vector WI; $^\ddag$ from axial WI. 
\vspace*{-.3cm}
\end{table*}
%___________________________________________________________________________
%________________________________________________________________
%________________________________________________________________
In this case, only the results obtained by using $m_K$ as experimental input 
have been considered.

The comparison among the results presented in table~\ref{tab:mslatt} becomes
more significant if one tries to correct them by taking into account for the 
dif\-fe\-rent systematic errors involved in the calculations. In this way, I 
obtain the estimates shown in fig.\ref{fig:msave} as empty circles. 

All results in table~\ref{tab:mslatt} have been obtained adopting a 
non-perturbative renormalization te\-chni\-que, with the only exceptions of the 
two CP-PACS determinations~\cite{mq_cppacs99,mq_cppacs00}, in which the quark 
mass renormalization constant has been evaluated using one-loop perturbation 
theory. An additional uncertainty, due to the use of perturbation theory, should
then be added to these results, to account for the corresponding systematic 
error. This uncertainty, however, is difficult to be quantified, because two 
different renormalization constants ($Z_S^{-1}$ and $Z_A/Z_P$) have been used in
the calculation and, moreover, non-perturbative determinations of 
renormalization constants with the Iwasaki action used in 
ref.~\cite{mq_cppacs00} have not been performed yet. In fig.~\ref{fig:msave}, a 
systematic error of 10\% has been added to the CP-PACS results. This error, 
however, may be underestimated, according to the analysis of the previous 
section (see table~\ref{tab:npr}).

The values of the strange quark mass by AL\-PHA-UKQCD~\cite{mq_alpha99} and 
QCDSF~\cite{mq_qcdsf99} have been obtained by fixing the lattice spacing from 
the value of the phenomenological parameter $r_0$~\cite{r0}, rather than from 
$m_\rho$ as in all other determinations. QCDSF finds that the choice of $r_0$ is
roughly equivalent to using $m_\rho$, while ALPHA-UKQCD finds that this is 
equivalent to setting the scale from $f_K$. ALPHA-UKQCD also estimates that a 
10\% higher value of $\ms$ would have been obtained by fixing the scale from the
value of the nucleon mass. Since the scale determined from $m_\rho$ turns out to
be typically in between the values obtained from $f_K$ and $m_N$, in order to
compare with the other determinations I have increased by 5\% in 
fig.\ref{fig:msave} the values of $\ms$ obtained by ALPHA-UKQCD and QCDSF.

The other corrections to the results presented in table~\ref{tab:mslatt} are
smaller. The APE determinations of refs.~\cite{mq_ape99,mq_ape98}, obtained with
the non-perturbative Clover action at $\beta=6.2$, have not been 
extra\-po\-la\-ted to the continuum limit. In this case, the more extensive 
analysis by ALPHA-UKQCD~\cite{mq_alpha99} shows that the value of the strange 
quark mass is underestimated by approximately 3\%. Finally, in the calculations 
by APE~\cite{mq_ape98} and JLQCD~\cite{mq_jlqcd99} the conversion from the 
non-perturbative RI-MOM renormalization scheme to the $\msb$ scheme has been 
done at the N$^2$LO in perturbation theory~\cite{francolub}, since at the time 
when these studies have been performed the N$^3$LO perturbative calculation 
of~\cite{chetyrkin} was not available yet. In fig.\ref{fig:msave}, in order to 
account for the difference between N$^2$LO and N$^3$LO, the results of 
refs.~\cite{mq_ape98} and \cite{mq_jlqcd99} have been decreased by 3\%.

I would like to emphasize that the previous discussion, besides exploiting the 
different sources and sizes of systematic errors, also shows that a rather good 
level of statistical and systematic accuracy has been achieved in the
determinations of light quark masses within the quenched approximation. Indeed,
the values of $\ms$ shown in fig.~\ref{fig:msave} are in remarkable agreement 
within each other, and even more when the differences in the systematics have
been taken into account. From the spread of the results, and accounting for the 
ty\-pi\-cal uncertainty on the scale in the quenched case ($\sim 10\%$), I 
obtain as a final average of the lattice results, within the quenched 
approximation,
\be
\ms(2\gev)^{\mbox{\scriptsize{\rm QUEN}}} = (110 \pm 15) \mev
\label{eq:msq}
\ee 
This value is shown in fig.\ref{fig:msave}, together with the se\-ve\-ral lattice
results and the average value of $\ms$ quoted by the PDG~\cite{pdg}. Note that 
the uncertainty in eq.~(\ref{eq:msq}) is approximately three times smaller than 
the one quoted by the PDG, although, in the former, the effect of the quenching 
approximation has not been taken into account yet. 

\begin{figure}[htb]
%\vspace*{-.55cm}
\epsfxsize=7.5cm 
\epsfbox{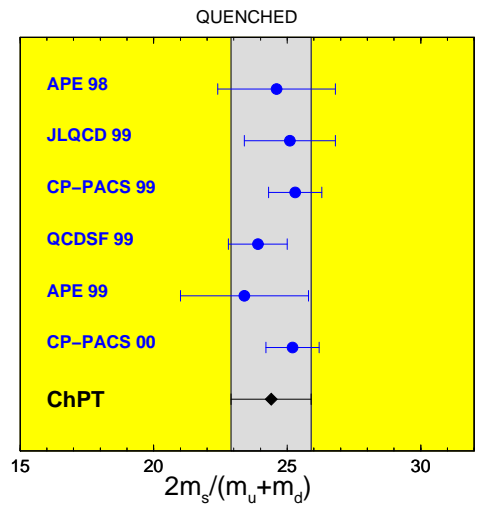}
\vspace*{-1.3cm}
\caption{\it Ratio of the strange to the average up-down quark masses as obtained
from recent lattice calculations in the quenched approximation (circles). The 
errors are my estimates based on the original data. The lowest point (diamond) 
represents the prediction of chiral perturbation theory.}
\label{fig:mudave}
\vspace*{-.55cm}
\end{figure}

In the light quark sector, the other quantity, besides $m_s$, which is 
accessible to lattice calculations is the average value of the up and down quark
masses, $m_{ud} \equiv\left(m_u + m_d\right)/2$. Rather then the value of 
$m_{ud}$ itself, however, I find more convenient to consider the ratio of the 
strange to the ave\-ra\-ge up-down quark masses, because in this ratio many of 
the systematic uncertainties, coming for instance from the renormalization 
constants or the choice of the scale, are expected to partially cancel. The 
values of $m_s/m_{ud}$ from the most recent lattice calculations are collected 
in table~\ref{tab:mudlatt} and shown in fig.~\ref{fig:mudave}.
%___________________________________________________________________________
\begin{table}[t]
\caption{Quenched lattice results for the ratio of the strange to the average 
up-down quark masses.}
\label{tab:mudlatt}
\begin{tabular}{lc}
\hline  & $2 m_s/(m_u+m_d)$ \\  \hline
APE 98 \cite{mq_ape98}            & $24.6 \pm 2.2$ \\
JLQCD 99 \cite{mq_jlqcd99}        & $25.1 \pm 1.7$ \\
CP-PACS 99 \cite{mq_cppacs99}     & $25.3 \pm 1.0$ \\
QCDSF 99 \cite{mq_qcdsf99}        & $23.9 \pm 1.1$ \\
APE 99 \cite{mq_ape99}            & $23.4 \pm 2.4$ \\
CP-PACS 00 \cite{mq_cppacs00}     & $25.2 \pm 1.0$ \\
\hline
\end{tabular}\\[2pt]
The errors are my estimates based on the original data.
\vspace*{-.3cm}
\end{table}
%___________________________________________________________________________
%________________________________________________________________
%________________________________________________________________
The quoted errors are my estimates based on the original data, because the 
value of this ratio has not been quoted directly in the original papers.

The relevant result, shown in fig.~\ref{fig:mudave}, is that the lattice 
predictions for $m_s/m_{ud}$ are in very good agreement with the value $24.4 
\pm 1.5$ predicted by chiral perturbation theory~\cite{leutwyler}. By using this
information and eq.~(\ref{eq:msq}), I obtain the quenched ave\-ra\-ge:
\be
\mud(2\gev)^{\mbox{\scriptsize{\rm QUEN}}} = (4.5 \pm 0.6) \mev \, .
\label{eq:mudq}
\ee 

Finally, I would like to mention that the first (quenched) lattice calculation 
of $\mud$ with overlap fermions has been presented at this conference~\cite{liu}.
The preliminary result is $\mud(2\gev)$ =5.5(5) MeV, in good agreement with the
ave\-ra\-ge~(\ref{eq:mudq}).
  
\subsection{Unquenched Results}
Several unquenched calculations of light quark masses, with two flavours of 
dynamical quarks, have been presented at this conference. A compilation of the 
results, together with details of the simulations, is given in 
table~\ref{tab:mqunq}. With the exception of the SESAM~\cite{mqu_sesam} and 
CP-PACS~\cite{mq_cppacs00} determinations, all other results presented in the 
table are still preliminary.

A striking result shown in table~\ref{tab:mqunq} is the value of the ratio of
the strange to the average up-down quark masses obtained by 
SESAM~\cite{mqu_sesam}: $m_s/m_{ud}\simeq 55$. This value is larger, by roughly 
a factor 2, than all other quenched and unquenched lattice determinations 
(including the quenched result by SESAM~\cite{sesam}), and in disagreement with 
the prediction of chiral perturbation theory. The reason for such a large value 
of $m_s/m_{ud}$ obtained by SESAM is the way in which the unquenched analysis 
has been performed. Indeed, as discussed below, by following a different 
procedure they obtain $m_s/m_{ud}\simeq 30$~\cite{sesam}, which is much closer 
to the expected value. I believe this point is worth to be discussed in details,
because it is also of relevance for other unquenched studies. 

In numerical simulations with dynamical fer\-mions all physical quantities, as
well as the lattice spacing and the critical value of the hopping parameter, are
functions of the two parameters entering in the action, namely the coupling
constant and the sea quark mass. Theoretical considerations suggest that the
dependence of the lattice spacing on the value of the sea quark mass should be 
exponential~\cite{mbunq}. Indeed, the lattice scale depends exponentially on the
effective coupling constant which, in turn, is affected by the values of quark
masses because quarks run in the loops. A large dependence of the lattice
spacing on the sea quark mass is in fact observed in actual calculations. For 
instance, JLQCD~\cite{mqu_jlqcd} finds that, in the range of $K_{sea}$ used in 
the simulation (corresponding to $m_{PS}/m_{V}\simeq 0.6-0.8$), the inverse
lattice spacing varies approximately by 30\%, between $\sim 1.5$ and $\sim 2.0$ 
GeV. The important point is that such a dependence, if not properly taken into 
account, may lead to uncontrolled systematic effects in the various
extrapolations, in both the sea and valence quark masses.

In order to control these effects, an appropriate procedure for the unquenched
analysis has been advocated~\cite{mbunq,ukunq}. It consists in performing all 
the analysis, including the calibration of the lattice spacing and the 
calculation of the relevant physical quantities, at each fixed value of 
$K_{sea}$. The simulations performed at fixed $K_{sea}$ can be thought of as 
{\em pseudo-quenched simulations}, which come closer to the description of the 
real world as the sea quark mass approaches its physical value. Only when all 
the quantities, computed at fixed $K_{sea}$, are expressed in physical units, 
the results can be extrapolated in the sea quark mass.
%___________________________________________________________________________
\begin{table*}
\caption{Simulation details and physical results of unquenched lattice
calculations of light quark masses.}
\label{tab:mqunq}
\renewcommand{\tabcolsep}{0.2pc} % enlarge column spacing
\renewcommand{\arraystretch}{1.2} % enlarge line spacing
\begin{tabular}{lcccclccc}
\hline  
& Action & $a^{-1}$ & $\#_{(\beta,K_{sea})}$ & $Z_m$ & 
\multicolumn{2}{c}{$\ \ms(2 \gev)$} & $m_s/m_{ud}$ & 
$\ms^{\mbox{\scriptsize{\rm QUEN}}}/\ms^{\mbox{\scriptsize{\rm UNQ}}}$
\\ \hline
SESAM 98 \cite{mqu_sesam}  & Wilson & $ 2.3 \gev$ & 4 &  PT  
      & $\, $ 151(30) &$(m_{K,\phi})$& 55(12) & 1.10(24)
      \\[8pt]
MILC 99 \cite{mqu_milc} & Fatlink & $ 1.9 \gev$ & 1 &  PT  
      & $\begin{array}{l} 113(11) \\ 125(9) \end{array}$
      & $\begin{array}{c} (m_K) \\ (m_\phi) \end{array}$ & 22(4) & 1.08(13)
      \\[16pt]
APE 00 \cite{mqu_ape} & Wilson & $ 2.6 \gev$ & 2 &  NP-RI  
      & $\begin{array}{l} 112(15) \\ 108(26) \end{array}$
      & $\begin{array}{c} (m_K) \\ (m_\phi) \end{array}$ & 26(2) & 1.09(20)
      \\[16pt]
CP-PACS 00 \cite{mq_cppacs00} & MF-Clover & $a \to 0$ & 12 & PT 
      & $\begin{array}{l} 88^{+4}_{-6} \\ 90^{+5}_{-11} \end{array}$
      & $\begin{array}{c} (m_K) \\ (m_\phi)\end{array}$ & 26(2) &1.25(7)
      \\[16pt]
JLQCD 00 \cite{mqu_jlqcd} & NP-Clover & $ 2.0 \gev$ & 5 & PT
      & $\begin{array}{l} 94(2)^\dag \\ 88(3)^\ddag \\ 109(4)^\dag \\ 
      102(6)^\ddag \end{array}$ 
      & $\begin{array}{c} (m_K) \\ \\ (m_\phi)\end{array}$ & --- & --- 
      \\[16pt] \hspace{-0.3truecm}
$\begin{array}{l} \mbox{\rm QCDSF +} \\ \mbox{\rm UKQCD 00 \cite{mqu_ukqcdsf}} 
      \end{array}$ & NP-Clover & $ 2.0 \gev$ & 6 &  PT & $\, $ 90(5) 
      & $(m_K)$ & 26(2) & --- \\
\hline
\end{tabular}\\[2pt]
$^\dag$ From vector WI; $^\ddag$ from axial WI. The errors on the ratios 
$m_s/m_{ud}$ and 
$\ms^{\mbox{\scriptsize{\rm QUEN}}}/\ms^{\mbox{\scriptsize{\rm UNQ}}}$ are my
estimates based on the original data.
\vspace*{-.3cm}
\end{table*}
%___________________________________________________________________________

\begin{figure}[htb]
%\vspace*{-.55cm}
\epsfxsize=7.5cm 
\epsfysize=7.0cm 
\epsfbox{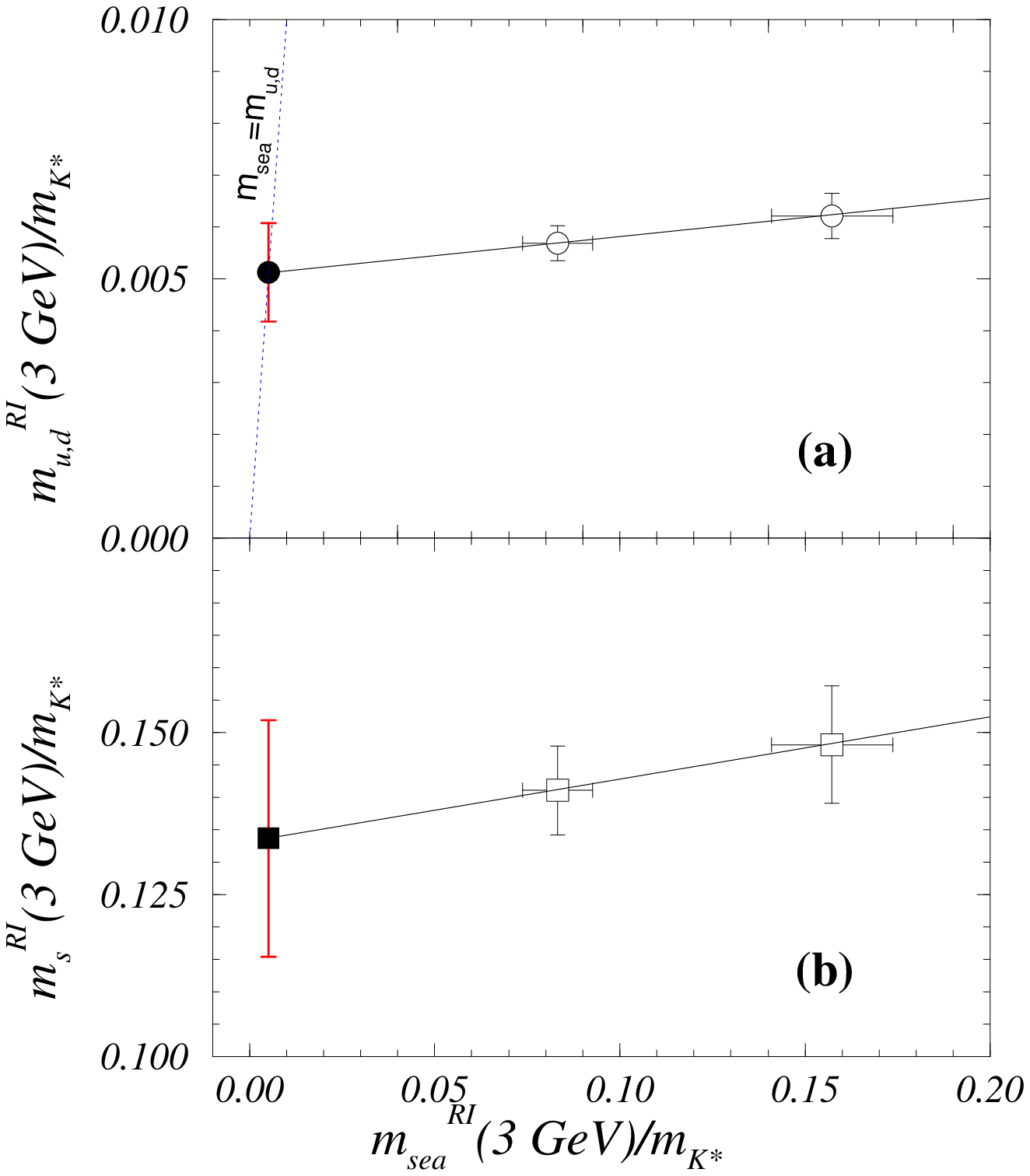}
\vspace*{-1.3cm}
\caption{\it Extrapolation of the average up-down (a) and the strange (b) quark
masses, computed at fixed $K_{sea}$, to the physical value of the sea quark 
mass, $m_{sea}=m_{ud}$.}
\label{fig:mqunq}
\vspace*{-.55cm}
\end{figure}

This procedure has been followed by APE~\cite{mqu_ape} in the calculation of the
strange quark mass. At two fixed masses of the sea quark, both the strange and 
the average up-down quark masses have been computed. The results are shown 
in fig.~\ref{fig:mqunq}, in terms of the  ratios $m_{ud}/m_{K^\ast}$ and $m_{s}/
m_{K^\ast}$. Presenting these ratios is equivalent to show the quark masses in 
physical units.
%________________________________________________________________
%________________________________________________________________
The values of $m_{ud}$ are then extrapolated in the sea quark mass, also
computed in physical units, up to the point corresponding to $m_{sea}=
m_{ud}$. Once the phy\-si\-cal value of $m_{ud}$ has been determined in this 
way, a similar extrapolation is performed for the strange mass. From 
fig.~\ref{fig:mqunq} one can see that the final extrapolations in the sea quark 
mass are indeed smooth.%
\footnote{Since, however, only two values of $K_{sea}$ have been simulated by
APE, the results should be taken as preliminary.}

In the analysis by SESAM and CP-PACS the pseudoscalar and vector meson masses 
have been fitted simultaneously in both the sea and the valence quark masses, 
with all quantities expressed in lattice units. The large 
dependence of the lattice spacing on the sea quark mass may not be taken 
properly into account in this way. Indeed, SESAM finds that the value of 
$m_s/m_{ud}$ changes by a factor 2 when the analysis, instead, is performed at 
fixed values of $K_{sea}$, as advocated before. Moreover, the result obtained 
with the analysis performed at fixed $K_{sea}$ is much closer to the expected 
value of this ratio. The systematic effects, introduced by the ne\-gle\-cting of
the lattice spacing dependence on the sea quark mass, are difficult to evaluate 
a priori, and depend on the specific values of the parameters used in the 
calculation. JLQCD, for instance, attempted the chiral extrapolations by using 
either dimensionless or dimensionful quantities~\cite{mqu_jlqcd} (although 
considering different orders in the expansions), and obtained for $m_s$ 
compatible results. 

Since unquenched studies of light quark masses have not yet reached the same 
degree of accuracy achieved in quenched calculations, in order to obtain an 
estimate of the quenching effect it is convenient to consider directly the ratio
between quenched and unquenched results. Some of the systematic errors are 
expected to cancel in this ratio, when both determinations are performed by 
using the same action, the same renormalization procedure and the same choice of
physical inputs. The results for the ratio 
$\ms^{\mbox{\scriptsize{\rm QUEN}}}/\ms^{\mbox{\scriptsize{\rm UNQ}}}$
are presented in table~\ref{tab:mqunq}. In the very extensive calculation 
performed by CP-PACS the value 
$m^{\mbox{\scriptsize{\rm QUEN}}}_s/m^{\mbox{\scriptsize{\rm UNQ}}}_s=1.25(8)$
is obtained, suggesting a sizable decrease of the quark mass in the unquenched 
case. This result, however, is not confirmed by the other (less extensive) 
studies by SESAM, APE and MILC, which find a decrease, in the unquenched case, 
rather of the order of 10\%. Given these results, and bearing in mind the
uncertainties in the analysis methods, I believe that, in order to quote a 
final estimate of light quark masses, it is more appropriate to include the
quenching error as an additional systematic uncertainty in the quenched averages
(eqs.~(\ref{eq:msq}) and (\ref{eq:mudq})), rather than varying the central 
values. Assuming this error to be of the order of 20 MeV in the case of the 
strange mass, I then obtain: 
\bea
&& \ms(2\gev) = (110 \pm 15 \pm 20) \mev \label{eq:ms} \\
&& \mud(2\gev) = (4.5 \pm 0.6 \pm 0.8) \mev \label{eq:mud}
\eea

Note also that in the unquenched case (see table~\ref{tab:mqunq}) the 
differences between the determinations of the strange quark mass from $m_K$ and 
$m_\phi$, if any, are rather small.

\section{THE $\mathbf{b}$-QUARK MASS}\label{sect:heavy}
The $b$-quark cannot be directly simulated on the lattice since its mass is 
larger than typical va\-lues of the ultraviolet cutoff in present lattice 
calculations ($a^{-1} \sim 3$ GeV). The $b$ mass, however, is also larger than 
the typical energy scale of strong interactions. Thus, the heavy degrees of 
freedom of the $b$-quark can be integrated out. Beauty hadrons may be therefore 
simulated on the lattice in the framework of a low-energy effective theory. 
Indeed, in the past years, many lattice simulations of Heavy Quark Effective 
Theory (HQET) and Non-Relativistic QCD (NRQCD) have been performed.

Within the effective theory, the mass of a $B$-meson, $M_B$, is related to the 
pole mass of the $b$-quark. Up to higher-order $1/m_b$ corrections, this 
relation has the form:
\be
M_B = m^{\mbox{\scriptsize{\rm pole}}}_b + \varepsilon - \delta m
\label{eq:mbpole}
\ee
where $\varepsilon$ is the so-called binding energy, which can be computed
non-perturbatively by a nu\-me\-ri\-cal simulation, and $\delta m$ is the 
residual mass, ge\-ne\-ra\-ted by radiative corrections, which can be 
eva\-lu\-a\-ted in
perturbation theory. The perturbative calculation of the residual mass and 
the non-perturbative (lattice) calculation of the binding energy thus allow a 
determination of the $b$ pole mass. The result can be then translated into the 
$\msb$ mass, $\mb$, by using perturbation theory. 

An important observation, concerning this procedure, is that the binding energy
$\varepsilon$ is not a phy\-si\-cal quantity, and it is affected by power 
divergences proportional to the inverse lattice spacing, $1/a$. These
divergences are canceled by similar singularities in the residual mass $\delta 
m$. Moreover, the pole mass $m^{\mbox{\scriptsize{\rm pole}}}_b$, and 
consequently $\delta m$, are also affected by renormalon singularities, which 
introduce in their definitions an uncertainty of the order of 
$\Lambda_{QCD}$~\cite{mpole1,mpole2}. These singularities are then canceled by 
the perturbative series relating the pole mass and the $\msb$ 
mass~\cite{renormalons}. As a result, the $\msb$ mass, $\mb$, is a finite, well 
defined, short-distance quantity. In the actual calculation, however, since the 
residual mass is computed up to a finite order in perturbation theory, only a 
partial cancellation of both power divergences and the renormalon singularities 
occurs. For this reason, it is crucial to compute the residual mass $\delta m$ 
up to the highest possible order in perturbation theory. Moreover, because of 
the incomplete cancellation of power divergences, the lattice spacing in such 
calculations cannot be taken too small.

At present, the most accurate determination of $\delta m$ has been obtained in
the framework of HQET. In terms of the lattice bare coupling constant, the 
perturbative expansion can be written as:
\be
a \, \delta m = C_1\, \alpha_0 + C_2 \, \alpha_0^2 + C_3 \, \alpha_0^3 + \dots
\ee
The two-loop coefficient, $C_2$ has been derived ana\-ly\-ti\-cal\-ly, by 
Martinelli and Sachrajda ~\cite{martisach}, from the study of the asymptotic 
behaviour of Wilson loops. In the quenched case ($N_f=0$), also the coefficient 
$C_3$ has been eva\-lua\-ted, by Di Renzo and Scorzato~\cite{direnzo}, using the 
method of numerical stochastic perturbation theory. An important check of this 
approach is obtained from the comparison of their numerical results for the 
first two coefficients, $C_1=2.09(4)$ and $C_2=10.7(7)$, with the analytically 
known values: $C_1=2.1173$ and $C_2=11.152$~\cite{martisach}. The value obtained
for the third coefficient, $C_3=86.2(5)$, is also consistent with the 
determination by Lepage {\em et al.}~\cite{trottier}, $C_3=81(2)$, obtained, 
within a completely different approach, by fitting the results of small coupling
Monte Carlo calculations. 

These combined theoretical efforts allow a determination of the $b$-quark mass 
which is accurate, in the quenched approximation, up to the N$^3$LO. Two 
independent results have been obtained so far:
\bea
\mb^{\mbox{\scriptsize{\rm \ QUEN}}} = \left( 4.30 \pm 0.05 \pm 0.05 \right) 
\gev && \hspace{-0.3truecm} \cite{gimenez} \label{eq:mbq1} \\
\mb^{\mbox{\scriptsize{\rm \ QUEN}}} = \left( 4.34 \pm 0.03 \pm 0.06 \right) 
\gev && \hspace{-0.3truecm} \cite{davies} \label{eq:mbq2}
\eea
nicely in agreement within each other. Re\-mar\-kably, the two determinations 
have been derived using different approaches. The APE result (\ref{eq:mbq1}) has
been obtained within HQET, while the determination (\ref{eq:mbq2}), by Collins
{\em et al.}, is obtained from the NRQCD study of the $B$-meson spectrum in the 
static limit. The last error in eqs.~(\ref{eq:mbq1}) and (\ref{eq:mbq2}) 
represents the residual uncertainty due to the neglecting of higher orders in 
the perturbative expansion of $\delta m$. For comparison, this uncertainty was 
estimated to be of the order of 200 MeV at NLO~\cite{mbnlo_gim} and 100 MeV at 
N$^2$LO~\cite{martisach} respectively.

Results compatible with those in eqs.~(\ref{eq:mbq1}) and (\ref{eq:mbq2}) have 
been obtained from the study of the $B$ and the $\Upsilon$ spectrum with 
NRQCD~\cite{nrqcd_B,nrqcd_Y}. In this case, however, only the coefficient $C_1$ 
is known, so that the achieved accuracy is at NLO. These results are thus 
affected by a larger perturbative uncertainty, which can be estimated, on the
basis of a renormalon analysis~\cite{renormalons}, to be of the order of 100 MeV.

The first unquenched calculation of the $b$-quark mass has been performed by APE
this year~\cite{mbunq}. The perturbative accuracy is at N$^2$LO, since the
three-loop coefficient, in the expansion of $\delta m$, is yet unknown in the 
unquenched case. Their result,
\be
\mb = \left( 4.26 \pm 0.06 \pm 0.07 \right) \gev \, .
\label{eq:mbunq}
\ee
represents to date the most accurate determination of the $b$-quark mass from 
lattice QCD calculations. Remarkably, the relative  uncertainty is reduced at 
the level of 2\% (90 MeV). 

A preliminary analysis~\cite{davies} performed in the static limit with NRQCD
indicates a decrease of $\mb$, in the unquenched case, by approximately 70 MeV.
Combined with the quenched determination (\ref{eq:mbq2}), this result leads to 
an estimate of the $b$-quark mass in good agreement with (\ref{eq:mbunq}).

\section{CONCLUSIONS}\label{sect:conclu}
Significant progresses in lattice calculations of light and heavy quark masses 
have been allowed by the introduction of improved actions and non-perturbative
renormalization techniques. 

Current determinations of light quark masses have reached a good level of 
statistical and systematic accuracy, which is of the order of 10\% within the 
quenched approximation. The main source of uncertainty, in this case, is due to 
the quenched approximation. Several unquenched results, with two flavours of 
dynamical quarks, have been presented at this conference. The effect of 
quenching is found to be in the range of 10-20\%, but further investigations 
are required to obtain a firmer estimate. 

I have not reported, in this review, lattice calculations of the charm quark
mass, since recent results for this quantity are missing. The charm mass is of 
great phenomenological interest since it enters, through the heavy quark
expansion, the theoretical predictions of several cross sections and decay
rates. For this reasons, new lattice calculations with the non-perturbatively 
improved action and non-perturbative re\-nor\-ma\-li\-za\-tion techniques is 
highly recommended.

Enormous progresses have been achieved in lattice calculations of the bottom 
quark mass. The present accuracy is at N$^3$LO in the quenched approximation, 
and at N$^2$LO in the unquenched case. The relative uncertainty is reduced at 
the impressive level of 2\%. 

Final averages of lattice calculations for quark masses have been given in 
eqs.~(\ref{eq:ms}), (\ref{eq:mud}) and (\ref{eq:mbunq}). By combining statistical 
and systematic errors, these correspond to:
\bea
&& \mud(2\gev) = (4.5 \pm 1.0) \mev \nonumber \\
&& \ms(2\gev) = (110 \pm 25) \mev \\
&& \mb(\mb) = ( 4.26 \pm 0.09) \gev \nonumber \, .
\eea

\section{ACKNOWLEDGEMENTS}
I thank D.~Becirevic, C.~Bernard, R.~Burkhalter, C.~Davies, F.~Di Renzo, 
V.~Gimenez, L.~Giu\-sti, T.~Kaneko, H.~Leutwyler, K.F.~Liu, G.~Martinelli, 
R.~Sommer, M.~Wingate and H.~Wittig for discussions and private communications.


\begin{thebibliography}{9}

\bibitem{alpha} K.~Jansen {\em et al.}, Phys. Lett. {\bf B372} (1996) 
275, {\tt hep-lat/9512009};\\
M.~L\"uscher {\em et al.}, Nucl. Phys. {\bf B478} (1996) 365, 
{\tt hep-lat/9605038};\\
M.~L\"uscher {\em et al.}, Nucl. Phys. {\bf B491} (1997) 323, 
{\tt hep-lat/9609035};\\
M.~L\"uscher {\em et al.}, Nucl. Phys. {\bf B491} (1997) 344, 
{\tt hep-lat/9611015}.

\bibitem{npm} G.~Martinelli {\em et al.}, Nucl. Phys. {\bf B445} (1995) 81, 
{\tt hep-lat/9411010}.

\bibitem{npm_sf} ALPHA Collaboration, S.~Capitani {\em et al.}, Nucl. Phys. 
{\bf B544} (1999) 669, {\tt hep-lat/\\9810063}. 

\bibitem{bochicchio} M.~Bochicchio {\em et al.}, Nucl. Phys. {\bf B262} (1985) 
331.

\bibitem{mq_ape99} 
APE Collaboration, D.~Becirevic {\em et al.}, Phys. Rev. {\bf D61} (2000)
114507, {\tt hep-lat/9909082}.

\bibitem{opeprop} K.~Lane, Phys. Rev. {\bf D10} (1974) 2605; \\
H.D.~Politzer, Nucl. Phys. {\bf B117} (1976) 397;\\ 
P.~Pascual and E.~de Rafael, Zeit. Phys. {\bf C12} (1982) 127.

\bibitem{qimp} G.~Martinelli {\em et al.}, EDINBURGH 96/28, unpublished;
C.~Dawson {\em et al.}, Nucl. Phys. Proc. Suppl. {\bf 63} (1998) 877,
{\tt hep-lat/\\9710027}.

\bibitem{bb_alpha} M.~Guagnelli {\em et al.}, DESY 00-131,
{\tt hep-lat/\\0009021}.

\bibitem{bb_lanl} T.~Bhattacharya {\em et al.}, LAUR-00-3538, \\
{\tt hep-lat/0009038}.

\bibitem{gupta} T.~Bhattacharya and R.~Gupta, Nucl. Phys. Proc. Suppl. {\bf 63}
(1998) 95, {\tt hep-lat/9710095}.

\bibitem{mq_alpha99}
ALPHA and UKQCD Collaboration, J.~Garden {\em et al.}, Nucl. Phys. {\bf B571}
(2000) 237, {\tt hep-lat/9906013}.

\bibitem{sint} S.~Sint, this proceedings.

\bibitem{mq_ape98} 
APE Collaboration, D.~Becirevic {\em et al.}, Phys. Lett. {\bf B444} (1998) 401,
{\tt hep-lat/9807046}. 

\bibitem{mq_jlqcd99} 
JLQCD Collaboration, S.~Aoki {\em et al.}, Phys. Rev. Lett. {\bf 82} (1999) 
4392, {\tt hep-lat/9901019}.

\bibitem{mq_rbc99} RBC Collaboration, M.~Wingate, Nucl. Phys. Proc. Suppl. 
{\bf 83} (2000) 221, {\tt hep-lat/\\9909101}.

\bibitem{alain} J.~Cudell, A.~Le Yaouanc and C.~Pittori, Phys. Lett. {\bf B454} 
(1999) 105, {\tt hep-lat/\\9810058}.  

\bibitem{lubicz} D.~Becirevic {\em et al.}, Nucl. Phys. Proc. Suppl. {\bf 83}
(2000) 863, {\tt hep-lat/9909039}.

\bibitem{giusti} L.~Giusti and A.~Vladikas, Phys. Lett. {\bf B488} (2000) 303,
{\tt hep-lat/0005026}.

\bibitem{n4lo1} K.G.~Chetyrkin, Phys.Lett. {\bf B404} (1997) 161,  
{\tt hep-ph/9703278}.

\bibitem{n4lo2} J.A.M.~Vermaseren, S.A.~Larin and T.~van Ritbergen,
Phys. Lett. {\bf B405} (1997) 327, {\tt hep-ph/9703284}.

\bibitem{allton} C.R.~Allton {\em et al.}, Nucl. Phys. {\bf B431} (1994) 667,
{\tt hep-ph/9406263}. 

\bibitem{sesam} SESAM Collaboration, N.~Eicker {\em et al.}, Phys. Lett. 
{\bf B407} (1997) 290, {\tt hep-lat/9704019}. 

\bibitem{pdg} Review of Particle Physics, {\it Eur. Phys. J.} {\bf C15}, 1 
(2000).                                                            

\bibitem{srbound} L.~Lellouch, E.~de Rafael and J.~Taron, Phys. Lett. {\bf B414}
(1997) 195, {\tt hep-ph/9707523}.

\bibitem{damir} D.~Becirevic, private communication.

%%%%%%%%%%%%%%%%%%%%   Light quark masses - Quenched    %%%%%%%%%%%%%%%%%%%%%

%\bibitem{mq_ape98} 
%APE Collaboration, D.~Becirevic {\em et al.}, Phys. Lett. {\bf B444} (1998) 
%401, {\tt hep-lat/9807046}. 

%\bibitem{mq_jlqcd99} 
%JLQCD Collaboration, S.~Aoki {\em et al.}, Phys. Rev. Lett. {\bf 82} (1999) 
%4392, {\tt hep-lat/9901019}.

\bibitem{mq_cppacs99} 
CP-PACS Collaboration, S.~Aoki {\em et al.}, Phys. Rev. Lett. {\bf 84} (2000)
238, {\tt hep-lat/9904012}.

%\bibitem{mq_alpha99}
%ALPHA and UKQCD Collaboration, J.~Garden {\em et al.}, Nucl. Phys. {\bf B571}
%(2000) 237, {\tt hep-lat/9906013}.

\bibitem{mq_qcdsf99} 
QCDSF Collaboration, M.~Gockeler {\em et al.}, Phys. Rev. {\bf D62} (2000) 
054504, {\tt hep-lat/\\9908005}.

%\bibitem{mq_ape99} 
%APE Collaboration, D.~Becirevic {\em et al.}, Phys. Rev. {\bf D61} (2000)
%114507, {\tt hep-lat/9909082}.

\bibitem{mq_cppacs00} 
CP-PACS Collaboration, A.~Ali Khan {\em et al.}, UTCCP-P-85, 
{\tt hep-lat/0004010}; see also R.~Burkhalter, this proceedings,
{\tt hep-lat/\\0010078}.

\bibitem{mq_rbc00} 
RBC Collaboration, M.~Wingate, this proceedings, {\tt hep-lat/0009023}.

%%%%%%%%%%%%%%%%%%%%%%%%%%%%%%%%%%%%%%%%%%%%%%%%%%%%%%%%%%%%%%%%%%%%%%%%%%%%%

\bibitem{francolub} E.~Franco and V.~Lubicz, Nucl. Phys. {\bf B531} (1998) 64, 
{\tt hep-ph/9803491}.

\bibitem{chetyrkin} K.G.~Chetyrkin and A.~Retey, Nucl. Phys. {\bf B583} (2000)
3, {\tt hep-ph/9910332}.

\bibitem{r0} R.~Sommer, Nucl.~Phys. {\bf B411} (1994) 839, {\tt
hep-lat/9310022}.
 
\bibitem{leutwyler} H.~Leutwyler, Phys. Lett. {\bf B378} (1996) 313, 
{\tt hep-ph/9602366}; see also H.~Leutwyler, this proceedings, 
{\tt hep-ph/0011049}. 

\bibitem{liu} S.J.~Dong {\em et al.}, {\tt hep-lat/0006004}; see also K.F.~Liu,
this proceedings, {\tt hep-lat/\\0011072}. 

%%%%%%%%%%%%%%%%%%%%   Light quark masses - Unquenched    %%%%%%%%%%%%%%%%%%%

\bibitem{mqu_sesam} SESAM Collaboration, N.~Eicker {\em et al.}, Phys. Rev. 
{\bf D59} (1999) 014509, {\tt hep-lat/9806027}. See also \cite{sesam}.

\bibitem{mqu_milc} MILC Collaboration, C.~Bernard, private communication.

\bibitem{mqu_ape} APE Collaboration, D.~Becirevic, V.~Gi\-me\-nez, L.~Giusti,
V.~Lubicz and G.~Martinelli, unpublished.

%\bibitem{mq_cppacs00} 
%CP-PACS Collaboration, A.~Ali Khan {\em et al.}, UTCCP-P-85, April 2000, 
%{\tt hep-lat/0004010}; R.~Burkhalter, this proceedings, {\tt hep-lat/0010078}.

\bibitem{mqu_jlqcd} JLQCD Collaboration, T.~Kaneko, this proceedings,
{\tt hep-lat/0010086}.

\bibitem{mqu_ukqcdsf} QCDSF and UKQCD Collaboration, D.~Ple\-i\-ter, this 
proceedings, {\tt hep-lat/0010063}.

%%%%%%%%%%%%%%%%%%%%%%%%%%%%%%%%%%%%%%%%%%%%%%%%%%%%%%%%%%%%%%%%%%%%%%%%%%%%%

\bibitem{mbunq} APE Collaboration, V.~Gimenez {\em et al.}, JHEP {\bf 0003} 
(2000) 018, {\tt hep-lat/0002007}.

\bibitem{ukunq} UKQCD Collaboration, C.R.~Allton {\em et al.}, Phys. Rev.
{\bf D60} (1999) 034507, {\tt hep-lat/\\9808016}.
 
\bibitem{mpole1} M.~Beneke and V.M.~Braun, Nucl. Phys. {\bf B426} (1994) 301,
{\tt hep-ph/9402364}.

\bibitem{mpole2} I.I.~Bigi {\em et al.}, Phys. Rev. {\bf D50} (1994) 2234,
{\tt hep-ph/9402360}. 

\bibitem{renormalons} G.~Martinelli and C.T.~Sachrajda, Nucl. Phys. {\bf B478}
(1996) 660, {\tt hep-ph/9605336}.

\bibitem{martisach} G.~Martinelli and C.T.~Sachrajda, Nucl. Phys. {\bf B559}
(1999) 429, {\tt hep-lat/9812001}. 

\bibitem{direnzo} F.~Di Renzo and L.~Scorzato, this proceedings, 
{\tt hep-lat/0010064}.

\bibitem{trottier} G.P.~Lepage {\em et al.}, Nucl. Phys. Proc. Suppl. {\bf 83}
(2000) 866, {\tt hep-lat/9910018}. 

\bibitem{gimenez} V.~Gimenez, private communication.

\bibitem{davies} C.~Davies, private communication. See also S.~Collins, {\tt
hep-lat/0009040}.

\bibitem{mbnlo_gim} V.~Gimenez, G.~Martinelli and C.T.~Sachrajda, Phys. Lett.
{\bf B393} (1997) 124, {\tt hep-lat/\\9607018}. 

\bibitem{nrqcd_B} A.~Ali Khan {\em et al.}, Phys. Rev. {\bf D62} (2000) 054505, 
{\tt hep-lat/9912034}.

\bibitem{nrqcd_Y} NRQCD Collaboration, C.T.H.~Davies, {\em et al.}, Nucl. Phys. 
Proc.Suppl. {\bf 73} (1999) 339, {\tt hep-lat/9809177}. 

\end{thebibliography}
\end{document}